\documentclass{mn2e}
\usepackage{natbib_jrm,times}
\usepackage{graphicx}
\bibpunct[, ]{(}{)}{;}{a}{}{,}

\def \aj {AJ}
\def \mnras {MNRAS}
\def \pasp {PASP}
\def \apj {ApJ}
\def \apjs {ApJS}

\def \aap {A\&A}
\def \nat {Nature}

\def \iaucirc {IAUC}
\def \aaps {A\&A Suppl.}

\begin{document}
\title[The Progenitor of SN 2005cs]{The Progenitor of SN 2005cs in the Whirlpool Galaxy}
\author[Maund, Smartt and Danziger]
{Justyn R.~Maund$^{1}$, Stephen J. ~Smartt$^{2}$ and I. John ~Danziger$^{3}$\\
$^{1}$Institute of Astronomy, University of Cambridge, Madingley Road, Cambridge CB3 0HA, England, UK\\
$^{2}$Department of Physics and Astronomy, Queen's University, Belfast BT7 1NN, Northern Ireland, UK\\
$^{3}$Osservatorio Astronomico di Trieste, INAF,  Via Tiepolo 11 34131 Trieste, Italia\\}
\maketitle
\begin{abstract}
The progenitor of SN 2005cs, in the galaxy M51, is identified in
pre-explosion HST ACS WFC imaging. Differential astrometry, with
post-explosion ACS HRC F555W images, permitted the identification of
the progenitor with an accuracy of $0.006\arcsec$.  The progenitor was
detected in the F814W pre-explosion image with $I=23.3\pm0.2$, but was
below the detection thresholds of the F435W and F555W images, with
$B<24.8$ and $V<25$ at $5\sigma$.  Limits were also placed on the $U$ and 
$R$ band fluxes of the progenitor from pre-explosion HST WFPC2 F336W and
F675W images. Deep images in the infra-red from NIRI on the Gemini-North 
telescope were taken 2 months prior to explosion, but the progenitor
is not clearly detected on these. The upper limits for the $JHK$
magnitudes of the progenitor were $J<21.9$,$H<21.1$ and $K<20.7$.  
Despite having a
detection in only one band, a restrictive spectral energy distribution
of the progenitor star can be constructed and a robust case is made
that the progenitor was a red supergiant with spectral type between
mid-K to late-M.  The spectral energy distribution
allows a region in the theoretical HR diagram to be determined which
must contain the progenitor star. The initial mass of the star is 
constrained to be  $M_{ZAMS}=9^{+3}_{-2}$M$_{\odot}$, which is 
very similar to the identified progenitor of the type II-P SN 2003gd, 
and also consistent with upper mass limits placed on five other similar
SNe. The upper limit in the deep K-band image is significant in that 
it allows us to rule out the possibility
that the progenitor was a significantly higher mass object enshrouded
in a dust cocoon before core-collapse. This is further evidence that the
trend for type II-P SNe to arise in low to moderate mass red supergiants
is real. 
\end{abstract}

\section{Introduction}
\label{intro}

Stellar evolution models predict that stars with initial masses
$>8M_{\odot}$ end their lives in a core-collapse induced supernova
(CCSN), and the CCSN mechanism is also related to the energetic
gamma-ray burst (GRB) phenomenon \citep{2003ApJ...599..394M}.  The
study of the progenitors of CCSNe is essential to both the
understanding of the evolution of SNe and the massive stars which produce
them. Fortuitous pre-explosion observations of the progenitor can allow 
the determination of the evolutionary state of the star prior to
SN explosion.  The progenitor of SN 1987A was found to be a blue
supergiant \citep{walb87a}, whereas the progenitor of SN 1993J was
identified as a red supergiant with a UV excess due to a hot binary companion
\citep{alder93j,maund93j}. In the cases of
SNe 1987A and 1993J the direct observations of the progenitors have
helped explain peculiarities in the SN explosions themselves. The
progenitor of the type II-P SN 2003gd was identified as a red
supergiant, the canonical prediction of stellar evolution models for
this type of SN
\citep{smartt03gd,2003PASP..115.1289V}. \citet{2005PASP..117..121L}
report the identification of a yellow supergiant progenitor for the type II-P SN 2004et.  Recently 
\citet{2005astro.ph..6436M} and \citet{2005astro.ph..1323M} have placed
limits on CCSNe, of various types, whose progenitors were not
detected on pre-explosion imaging.  \citet{2005astro.ph..1323M}
demonstrate the importance of high resolution post-explosion imaging
of SNe, using differential astrometry to locate (with sub-pixel
accuracy) the progenitor on pre-explosion images.

 In this letter we
report the identification of the progenitor of the Type II-P SN
2005cs in M51, the ``Whirlpool Galaxy.''  SN 2005cs was discovered by
\citet{2005IAUC.8553....1K} on 2005 June 28.9.  The SN was
spectroscopically classified as a Type II by
\citet{2005IAUC.8555....1M}, with P-Cygni profiles of the Balmer and
He lines.  \citet{2005IAUC.8555....2R} and \citet{2005IAUC.8556....2L}
reported the identification of two different stars as possible
progenitors for this object: a blue and a red star respectively.
\citet{2005IAUC.8556....2L} give the position of SN 2005cs as
$\mathrm{\alpha_{2000}=13^{h}29^{m}52.^{s}76}$,$\mathrm{\delta_{2000}=47\degr10\arcmin36.\arcsec11}$.  This implies an offset between SN 2005cs and the centre of M51 of
$\Delta\alpha=6\arcsec$ and $\Delta\delta=-64.7\arcsec$, or 2.7kpc at a distance of 8.4Mpc (\citealt{1997ApJ...479..231F}, with  P.A. and inclination of M51 as quoted by HyperLEDA\footnote{$\mathrm{http://www-obs.univ-lyon1.fr/hypercat/}$} as $163\degr$ and $47.5\degr$ respectively).
While this paper was in the process of submission, a similar study was submitted for publication by \citet{2005astro.ph..7394L}. There are some differences in the results of the two studies and we briefly highlight them in this letter where appropriate.

\section{Observations and Data Reduction}
\label{obs}
\begin{table}
\caption{\label{obstab} Pre-and post-explosion observations of the site of SN 2005cs
used in this paper.}
\begin{tabular}{lrrc}
\hline\hline
Date  & Instrument & Filter  & Exp. Time \\
      &            &         & (s)       \\
\hline
\multicolumn{4}{c}{{\bf Pre-explosion images}} \\
2005 Jan 20-21 & ACS/WFC& F435W & 2720\\
 & ACS/WFC & F555W & 1360 \\
 & ACS/WFC & F814W & 1360 \\
1999 Jul 21 & WFPC2    & F336W  & 1200 \\
& WFPC2    & F555W  & 1200 \\
            & WFPC2    & F675W  & 500\\
2005 Apr 29  & NIRI & $J$ (G0202) & 500 \\
               & NIRI & $H$ (G0203) & 500 \\
               & NIRI & $K$ (G0204) & 625 \\
~\\\hline
\multicolumn{4}{c}{{\bf Post-explosion images}} \\
  2005 Jul 24          & ACS/HRC & F555W & 80/160/1704 \\
\hline\hline
\end{tabular}
\end{table}

The pre- and post-explosion images analysed here are summarised in
Table \ref{obstab}.  Pre-explosion images of the
site of SN 2005cs were available in the Hubble Space Telescope (HST)
archive\footnote{http://archive.stsci.edu/hst/}, with multi-colour
imaging with the Wide Field Planetary Camera 2 (WFPC2) and Advanced
Camera for Surveys (ACS) Wide Field Channel (WFC).  Individual HST
WFPC2 observations were retrieved from the STScI archive.  These
observations were calibrated with the On-the-fly-recalibration (OTFR)
pipeline.  The site of SN 2005cs was located on the WF2 chip
in the 1999 WFPC2 observations, 
which has $\mathrm{0.1\arcsec~pix^{-1}}$.  
The WFPC2 observations
were combined for the extraction of cosmic rays and corrected for
geometric distortions.  Aperture photometry was then conducted on
these images using {\sc DAOphot}.  PSF photometry was conducted on the
WFPC2 observations using the {\sc HSTphot} package
\citep{dolphhstphot}.  {\sc HSTphot} includes 
corrections for chip-to-chip variations, charge transfer efficiency
and transformations from the WFPC2 photometric system to the
standard Johnson-Cousins magnitude system.  The WFPC2 pre-explosion
observation complemented the ACS pre-explosion imaging with
observations in F336W and F675W (approximately U and R
respectively).

The ACS pre-explosion images were acquired as part of
program GO-10452 (PI: S.V.W. Beckwith), which conducted a large deep
mosaic of M51 (NGC 5194) and its companion NGC 5195.  These
observations were in four bands F435W, F555W, F814W and F658N.  The
drizzled combined mosaic frames were available from
STScI \citep{2005BAAS},with scale $\mathrm{0.05\arcsec\;pix^{-1}}$.  Aperture and PSF
photometry was conducted on the mosaic frames using the {\sc PyRAF}
implementation of the {\sc DAOPhot} algorithm.  Empirical aperture
corrections were applied to the data, with the charge transfer
efficiency corrections of \citet{riesscte} and the colour transformation
equations and updated zeropoints of
\citet{acscoltran}.

Infrared pre-explosion observations of the site
of SN 2005cs were conducted with the Gemini North Telescope Near
Infra-red Imager (NIRI) in the $JHK$ bands.  These
observations were conducted as part of program GN-2005A-Q-49 (PI:
S. Smartt).  The f/6 camera of NIRI was used to provide imaging with
scale $\mathrm{0.117\arcsec\;pix^{-1}}$.  The observations were split
into a series of short sub-exposures, to overcome the sky background,
with small offsets between each sub-exposure.  On completion of the
series of observations on target, a comparable series of short
exposures at a large offset was immediately acquired to measure the
levels of the IR sky background.  The NIRI images were reduced and
combined, with sky subtraction, using the {\sc Iraf} {\it gemini}
package of reduction tools for the NIRI instrument.  Photometry on the
$JHK$ images was done by PSF fitting within the IRAF implementation of
DAOPHOT. The zero-point of the NIR filters was
determined using two standard stars (UKIRT Faint Standards 133 and
136) taken before and after the science frames. They gave comparable
results, and suggested that the period of the observations was stable
and photometric. The image quality was $0.5\arcsec$ in $K$ and $0.6\arcsec$
in $J$ and $H$. 

Post-explosion ACS High Resolution Channel (HRC) imaging was conducted as
part of program GO-10498 (PI: S. Smartt).  These observations were only acquired in the F555W, as good colour information on the surrounding stellar population was possible from the pre-explosion observations.  These were photometered in a
similar way to the pre-explosion ACS/WFC observation using {\sc
PyRAF} {\sc DAOPhot}.  The same corrections were applied, although
the magnitudes were left in the ACS VegaMag photometric system.

The geometric transformation between the F555W images of WFPC2 1999 observations with 2005
ACS/WFC observations was calculated, using the {\sc Iraf} task {\it
geomap}, with 11 common stars.  X- and Y- offsets
between the F555W image and other filter images from the same
observing run were calculated by cross-correlating the images, using
the {\sc Stsdas} {\it crossdriz} task with the offset calculated from
the cross correlation image using {\it shiftfind}.  The geometric
transformation was also calculated between the 2005 ACS/WFC F814W image and
the NIRI $J$-band image, with the shift calculated between the ACS/WFC
F555W and F814W images applied to bring the transformed NIRI $J$-band 
image to the same coordinate system as the ACS/WFC F555W image.  Similarly to the
HST pre-explosion observations, offsets between the $J$-band image and
the $H$ and $K$ images were calculated using the cross correlation method.  
The uncertainty in transforming the position of a star from one image 
to another was calculated as the sum in quadrature of the positional 
uncertainties in the
reference image and transformed image and the r.m.s. uncertainty of
the transformation, by {\it geomap}.  The
position uncertainties on the individual images were assessed by
measuring the scatter in the positions of the stars, measured using
the four centering algorithms of {\sc DAOPhot}, of all the stars used
to calculate the geometric transformation.  The individual image
position uncertainty was taken as the sum in quadrature of the mean
scatter and standard deviation about this mean in both the X and Y
directions.

\section{Observational Results}
\label{results}

SN 2005cs was detected in the post-explosion ACS/HRC F555W imaging
with $\mathrm{m_{F555W}=15.28\pm0.01}$.  The position of SN 2005cs on
the post-explosion frame was calculated to within $\pm0.002\arcsec$.
The position of the SN on pre- and post-explosion images in shown on 
Fig. 1.
The SN position on the pre-explosion ACS/WFC F555W image was
identified to within $0.006\arcsec$ (which is the combined
uncertainties on the determination of positions on the pre- and
post-explosion images and the geometric transformation). A progenitor
star is significantly detected on the F814W image with
$\mathrm{m_{F814W}=23.26\pm0.03}$.  There is no object in the
pre-explosion F435W and F555W images, and we estimate $5\sigma$
detection limits of $\mathrm{m_{F435W}}=24.9$ and
$\mathrm{m_{F555W}=25}$ (using the methods of Maund \& Smartt
2005). In order to add further limits to the spectral energy distribution
(SED) of the progenitor, we determined the limiting magnitude of the WFPC2 F336W and F675W images. Due to the lower resolution of these images
compared to the ACS, the progenitor position was too close to the 
brighter object to the southwest to allow accurate simultaneous PSF fitting. 
Hence a large (6 pixel) aperture flux was determined
at the progenitor position, and the flux of stars which had well determined
individual small aperture photometry was subtracted. This suggested there
was no residual flux from any object brighter than $\mathrm{m_{F336W}=21.1}$ and $\mathrm{m_{F675W}=23.6}$.
\citet{2005astro.ph..7394L} report a magnitude of $I=24.15\pm0.20$ for the 
progenitor which is significantly fainter than our measurement. We have
extensively checked our ACS photometry and zeropoints using several methods 
(e.g. DOLphot\footnote{http://purcell.as.arizona.edu/dolphot/}
 and several alternative experiments with PSF and aperture photometry)
and also compared the F814W magnitudes (from HSTphot and our own estimates)
of faint isolated stars in WFPC2 archive images with our ACS measured magnitudes.
We find no problems or inconsistencies in our analysis and confirm the 
magnitude as $\mathrm{m_{F814W}=23.26\pm0.03}$.

The location of the SN on the
pre-explosion Gemini NIRI imaging was identified to within
$0.078\arcsec$. There is a $JHK$ source clearly detected close to the
position of the $I$-band progenitor, however its position is more than
2$\sigma$ away and hence is not spatially coincident with the progenitor.
This object has also been detected by \citet{2005astro.ph..7394L} in 
NICMOS pre-explosion images, and is coincident with the $I$-band
star to the north west of the progenitor. At the position of the
progenitor in the $JHK$ NIRI images, the flux does appear somewhat
higher than the surrounding background, but given the ground-based
resolution there is no evidence for detection of a single source
(see Fig. 1). 
A further check was performed to determine that no object could be 
detected in the $K$-band. 
The bright objects on the image were subtracted off by PSF fitting, 
and fainter objects were identified on the resultant
frame. After three iterations of PSF
fitting and subtraction,  a boxcar median filtered image was
constructed, which should be representative of the varying galaxy
background. This was subtracted off the original frame and the
photometric procedure was repeated. There was no evidence of a single
point source at the progenitor position. 
The detection limits in the $JHK$-band images where
determined at the SN position. In the first instance statistical 
calculations of the 3$\sigma$ detection limits were determined from the 
detector characteristics and the sky noise. This was compared
with the measured magnitudes of the faintest real objects 
in the images which could be visually identified and which were
both well fit with a PSF and subtracted off cleanly. The 3$\sigma$
calculated limits were significantly fainter (by $\sim$0.7 mag)
than these objects, but the 
5$\sigma$ limits agreed well with the magnitudes of the faintest 
sources measured in the image. Hence we adopted the 
5$\sigma$ limits, which are:
$J=21.9$, $H=21.1$ and $K=20.7$.
The $JH$ limits are consistent with the NICMOS limits reported by 
\citet{2005astro.ph..7394L}, although the additional $K$-band upper limit we present has significant implications (see Section 4.3). 

\begin{figure*}
\label{diffimg}
\includegraphics{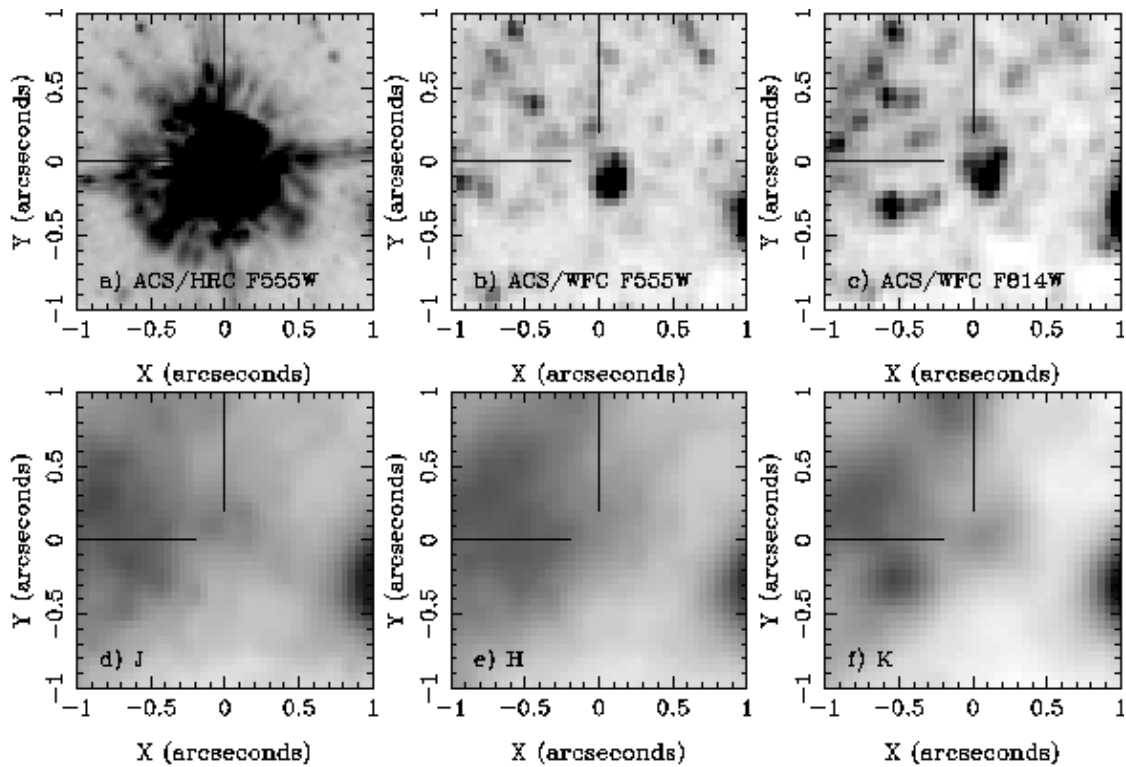}
\caption{Pre- and post-explosion images of the site of SN 2005cs in
M51. a) Post-explosion ACS/HRC F555W image (160 s) of SN 2005cs, the cross
hairs indicate the centre of the PSF. The observation was split into 
short exposures to avoid saturation of the SN centroid. 
b) Pre-explosion ACS/WFC F555W
image, in which the progenitor is not detected to $\mathrm{m_{F555W}=25}$ at $5\sigma$. c) Pre-explosion ACS/WFC F814W image, the progenitor is detected in this
image, indicated by the cross hairs, with $\mathrm{m_{F814W}=23.26}$.  d)
Pre-explosion NIRI $J$-band image.  e) Pre-explosion NIRI
$H$-band image.  f) Pre-explosion NIRI $K$-band image.  The images are
aligned to the reference ACS/WFC F555W image (panel b), such that
North is up and East is left.}
\end{figure*}

\section{Discussion}
\label{disc}
\subsection{Estimates of metallicity of the progenitor region}
\label{meta}
HII regions around M51 were spectroscopically studied by
\citet{2004ApJ...615..228B} for the purposes of abundance analysis.
The sight line corrected angular distance of SN 2005cs is $66.5\arcsec$ or 0.21$R_{0}$ (where $R_{0}$ is the
characteristic radius which \citet{2004ApJ...615..228B} give as
$5.4\arcmin$).  The oxygen abundance, at the radius of SN 2005cs, was
calculated to be $\mathrm{12+log(O/H)=8.66\pm0.11}$, which is 
similar to abundance in the H\,{\sc ii} region CCM\,56 ($>$8.58\,dex) closest to SN 2005cs.  \citet{2004ApJ...615..228B} point out that their new 
abundance determinations, which are based on directly determined 
electron temperatures, are significantly lower than previous estimates
in M51 and that the previously reported very metal rich abundances are 
likely to have been inaccurate. 
As \citet{2004A&A...417..751A} give the solar oxygen abundance
as $8.66\pm0.05$, this implies that the metallicity appropriate for
SN 2005cs and its progenitor star is very likely to be in the range 
$1\pm0.3Z_{\odot}$ 

\subsection{Estimates of Reddening}
\label{red}
\citet{2005IAUC.8555....1M} reported the presence of Galactic and
host galaxy components of the NaID lines in an early spectrum of SN
2005cs.  They reported that the equivalent width of the host-galaxy
component was similar in strength to the Galactic component of
0.2\AA.  \citet{2003fthp.conf..200T} provide relationships between
the strength of the NaID absorption and the reddening from the intervening
medium.  If one assumes that the strengths of the two NaID components may be added
then the total NaID line strength implies a reddening between
$E(B-V)=0.05-0.16$. \citet{2004ApJ...615..228B} record
the reddening coefficient $\mathrm{c(H\beta)}$ for the nearby HII region
CCM 56 of 0.23 which corresponds to $E(B-V)=0.16$ with a Galactic
\citet{1983MNRAS.203..301H} reddening law.
The three colour photometry of
red supergiants ($B-V > 0.2$) within $2\arcsec$ of SN 2005cs was also
used to estimate the reddening as detailed in \citet{2005astro.ph..1323M}. 
Stellar positions were plotted on a
two-colour diagram ($B-V$ vs. $V-I$) and compared with a theoretical supergiant
colour sequence.  The weighted mean displacement of the stars'
positions, along the reddening vector, from the theoretical colour
sequence provided a measure of reddening. This technique yielded a
value of $E(B-V)=0.12\pm0.01$, and we hence adopt a reddening of $E(B-V)=0.14\pm0.02$ throughout the rest of this paper.  

\subsection{The Progenitor of SN 2005cs}
There is a clear and unambiguous detection of the progenitor of 
SN2005cs in the ACS F814W image. As we have only one detection, the 
colour of the star makes the transformation from the F814W to standard
Johnson $I$ a little uncertain, and indeed all the transformations
between the ACS filters and standard $BVRI$ are subject to a similar
uncertainty. However from the non-detections in the other bands, we
can confidently say that it was a red star, later than mid K-type
and the corrections between the ACS and Johnson $VRI$ system are at most
0.2 mag, if the star was at the extreme M5 cool end
\citep{2005astro.ph..1323M}. Hence in the 
rest of this discussion we interpret the ACS filter magnitudes as
standard $BVRI$, and the relatively small uncertainties are included in the error
analysis. The non-detection of a 
source in the $BVR$-band images is a very robust argument that 
the star was red. The spectral energy distribution of the progenitor star 
in $BVRIJHK$ is shown in Fig. 2. This is consistent with a
red supergiant which is between K5 and M5. The star cannot have had bluer
colours than a mid K-type, or it would have been detected in the $VR$-bands. 

\begin{figure}
\label{sed}
\includegraphics[width=6cm,angle=270]{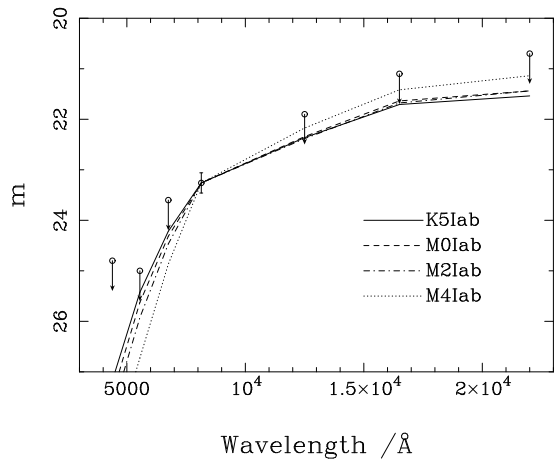}
\caption{The spectral energy distribution of the progenitor is constrained by the
$I$-band detection, and the upper limits in $VRJHK$.The star cannot have been 
bluer than a mid K-type, or it would have been detected in the $V$ and $R$ bands.
The colours of the K5Ia-M4Iab Galactic supergiants were taken from \citet{1985ApJS...57...91E}, 
and scaled to $I$=23.26 with appropriate reddening as discussed in the text. }
\end{figure}

Although there is a range in possible spectral types that would
fit the stellar SED, we can still set restrictive luminosity and
mass limits on the red progenitor star. Using the standard
reddening law of \citet{ccm89}, it has a $(V-I)_{0}>1.52$,
which corresponds to a supergiant redder than approximately K3
\citep{1985ApJS...57...91E}. The effective temperatures of red
supergiants, $(V-I)_{0}$ colours and bolometric corrections from 
\citet{1985ApJS...57...91E} and also \citet{drill00}
can then be used to determine a location box in the theoretical 
HR-diagram. This is shown in Fig. 3, with the stellar
evolution tracks from the Geneva group plotted at 
$Z=Z_{\odot}$ \citep{1994A&AS..103...97M}. 
The range in $\log L$ at a constant $T_{\rm eff}$ comes from 
the quadrature
combined errors in distance, reddening, measured $I$-band magnitude 
(including colour correction),
and bolometric correction (assuming plus or minus one spectral subtype). 
We can place quite a restrictive box on the location of the red supergiant
progenitor  in the HRD, even with a detection in only 
one band. The star was likely to have been a red supergiant of
initial mass 9$^{+3}_{-2}$M$_{\odot}$. 
\citet{2005astro.ph..7394L} report an initial mass of $7-9$M$_{\odot}$, 
and indeed the position of the progenitor is closer to 7M$_{\odot}$
in their figure. The 
difference of 1 magnitude in the $I$-band is responsible for the 
higher mass that we report, although within the errors the two studies
are in agreement.  This is a very similar mass
to that derived for the progenitor of SN~2003gd, which was also
a red supergiant of initial mass 8$^{+4}_{-2}$M$_{\odot}$ 
\citep{smartt03gd}. Both these supernovae were of type II-P 
\citep{2005astro.ph..7394L,2005MNRAS.359..906H} and future comparison
of the photometric and spectral evolution of SN~2005cs will determine
how similar they actually are. As pointed out by \citet{smartt03gd} and 
\citet{2005astro.ph..7394L}, there is growing evidence that SNe II-P
tend to arise in red supergiants with masses lower than $\sim$15M$_{\odot}$. 
There are restrictive mass limits, or detections,  now set for seven SNe II-P
(SNe 1999br, 1999em, 1999gi, 2001du, 2003gd, 2004et, 2005cs) and none of them 
are higher than 15M$_{\odot}$ 
\citep{2005astro.ph..1323M,smartt01du, 2005PASP..117..121L, 2003PASP..115..448V,
2003PASP..115.1289V}. One possibility often suggested to explain this is that the progenitors may be dust enshrouded red supergiants and hence appear to have lower mass and 
lower luminosity. Although the SNe themselves are not reddened, a plausible 
scenario is the destruction of this dust by the explosion 
\citep{1986MNRAS.221..789G}. However the deep $K$-band image argues against this scenario.
Even if the visual extinction were as high as $A_V=5$, the $K-$band images would be sensitive
to stars with $M_K<-9.5$, and with the bolometric corrections of \citet{1985ApJS...57...91E} this implies $\log L/L_{\odot} < 4.6$. Even when conservative errors on the distance, limiting
magnitudes and bolometric correction are included (resulting in $\pm$0.2\,dex), 
Fig.\,3 shows that this rules out red supergiants with masses greater than 12M$_{\odot}$. 

In summary, this letter presents the unambiguous detection of a red
supergiant as the progenitor of SN~2005cs. We show that the likely
initial mass for the star was in the range
$M_{ZAMS}=9^{+3}_{-2}$M$_{\odot}$. This adds to the emerging argument
that all type II-P supernovae lie in the low to moderate mass range of
red supergiants.  The upper limit from the deep $K$-band image
is evidence that these are not dust enshrouded higher mass
objects that have their cocoons destroyed in the explosion.

\begin{figure}
\label{hrd}
\rotatebox{-90}{
\includegraphics[width=6.5cm]{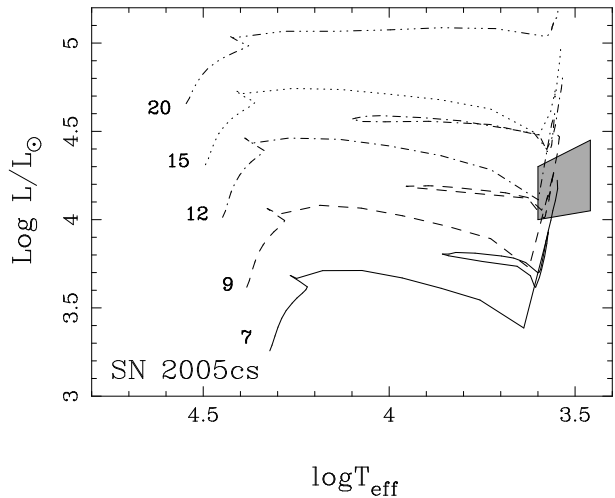}
}
\caption{The position of the progenitor of SN2005cs is restricted to lie in the grey
shaded region. The evolutionary tracks are the non-rotating models from the Geneva group
(Meynet et al. 1994), where the initial mass, in units of $M_{\odot}$, is indicated for each track. The cool limit of the allowed region is the lowest
observed $T_{\rm eff}$ of late M-type supergiants.}
\end{figure}

\section*{Acknowledgments}
Based  on observations made with the NASA/ESA Hubble Space
Telescope, and on observations with the  Gemini
Observatory, which is operated by the Association of Universities for
Research in Astronomy, Inc., under a cooperative agreement with the
NSF on behalf of the Gemini partnership. JRM and SJS
acknowledge financial support from PPARC.  We thank the referee, S. Ryder, for his useful comments and discussion.  

\bibliographystyle{mn2e}

\end{document}